\documentclass[5p,twocolumn]{elsarticle}
\pdfoutput=1
\usepackage{graphicx,epsfig}
\usepackage{amsmath,mathrsfs,amsfonts}
\usepackage {amssymb}
\usepackage {longtable}
\usepackage{multirow}
 \usepackage{enumerate}

\usepackage{bm}
\usepackage{amsfonts}
\usepackage{subfigure}
\usepackage{color}
\usepackage{relsize}
 \usepackage[utf8]{inputenc}
\usepackage{hyperref}

\newcommand{\avg}[1]{\langle #1\rangle}

\newcommand{\be}{\begin{equation}}
\newcommand{\ee}{\end{equation}}
\newcommand{\bea}{\begin{eqnarray}}
\newcommand{\eea}{\end{eqnarray}}

\biboptions{sort&compress}

\begin{document}

\begin{frontmatter}

\title{Effective dark energy through spin-gravity coupling }


\author[1]{Giovanni Otalora\fnref{myfootnote}}
\author[2,3,4]{Emmanuel N. Saridakis\fnref{myfootnote2}}
\address[1]{Departamento de F\'isica, Facultad de Ciencias, Universidad de Tarapac\'a, Casilla 7-D, Arica, Chile}
\address[2]{National Observatory of Athens, Lofos Nymfon, 11852 Athens, Greece}
\address[3]{CAS Key Laboratory for Research in Galaxies and Cosmology,  
 University of Science and Technology of China, Hefei, Anhui 230026,
China}
\address[4]{Departamento de Matem\'{a}ticas, Universidad Cat\'{o}lica del Norte, 
Avda.
Angamos 0610, Casilla 1280 Antofagasta, Chile}

\fntext[myfootnote]{giovanni.otalora@academicos.uta.cl}
\fntext[myfootnote2]{msaridak@noa.gr}

\begin{abstract}
We investigate cosmological scenarios with spin-gravity coupling. In 
particular, due to the spin of the baryonic and dark matter particles and its 
coupling to gravity, they probe an effective  spin-dependent metric, which 
can be calculated semi-classically in the   
Mathisson-Papapetrou-Tulczyjew-Dixon formalism. 
Hence, the usual field equations give rise 
to modified Friedmann equations, in which the extra terms can be identified as 
an effective dark-energy sector. Additionally, we obtain  an effective 
interaction between the matter and dark-energy sectors. In the case where the   
spin-gravity coupling switches off, we recover standard $\Lambda$CDM 
cosmology. We perform a dynamical system analysis  and we find a 
matter-dominated point that can describe the matter era, and a stable late-time 
solution  corresponding to acceleration and dark-energy domination. For small 
values of the 
 spin coupling  parameter, 
 deviations from $\Lambda$CDM concordance scenario are small, 
however for larger   values they can be brought to the desired 
amount, leading to different dark-energy equation-of-state parameter behavior, 
as well as to different transition redshift from acceleration to deceleration.
 Finally, we confront the model predictions with Hubble 
function data.

\end{abstract}

\end{frontmatter}


\section{Introduction}\label{Introduction}

Modified gravity is one of the two ways one may follow in order to explain 
universe acceleration \cite{CANTATA:2021ktz,Nojiri:2017ncd} (the other 
one being the introduction of the dark-energy sector 
\cite{Copeland:2006wr,Cai:2009zp}), while it has 
been also proved  to  be efficient in alleviating  the two  famous tensions of 
$\Lambda$CDM
cosmology, namely the $H_{0}$ and the $\sigma_{8}$ ones \cite{Abdalla:2022yfr}. 
However, a crucial additional advantage of modified gravity is that 
it may have improved renormalizability and thus is closer to a quantum 
description    \cite{Addazi:2021xuf}.  

One possibility is   the addition of quantum corrections to the action
\cite{Clifton:2011jh, Tsujikawa:2010zza, Nojiri:2006ri}. Other approaches to 
quantum gravity   consider that the standard energy-momentum dispersion 
relation is deformed near the Planck scale, since this may arise from  string 
field theory \cite{Kostelecky:1988zi}, loop quantum gravity 
\cite{Gambini:1998it}, and non-commutative geometry \cite{Carroll:2001ws}. 
Furthermore, in the context of ``doubly general relativity''   
\cite{Magueijo:2002xx,Amelino-Camelia:2000cpa,Amelino-Camelia:2000stu, 
Amelino-Camelia:2003xax} the modification of the dispersion relation 
leads to an effective spacetime metric which depends on the energy and momentum 
of the probe particle. Since   spacetime is represented by a one-parameter 
family of metrics, parametrized by the energy of the probe particle, this 
semiclassical approach is called Gravity's Rainbow \cite{Magueijo:2002xx}. 
Applications of Gravity's Rainbow to dark energy and inflation can be found in 
 \cite{Amelino-Camelia:2013gna,Ling:2006az,Garattini:2014rwa,Chatrabhuti:2015mws,
Waeming:2020rir, Leyva:2021fuo, Leyva:2022zhz}. 

However, interestingly enough,  an effective spacetime metric can  
alternatively arise in the context of a semiclassical description of the 
spinning particle in an arbitrary gravitational field 
\cite{Hehl:1974cn,Hehl:1976kj, Gasperini:1986mv,Deriglazov:2015bqa},
without the need of a modified   dispersion relation.
 For instance,  in   \cite{Deriglazov:2015bqa} (see also Refs. 
\cite{Deriglazov:2014zzm,GuzmanRamirez:2013ynp}) the authors have constructed a 
Lagrangian formulation for the  Mathisson-Papapetrou-Tulczyjew-Dixon (MPTD)
equations \cite{Mathisson:1937zz, Papapetrou:1951pa, Tulczyjew,Dixon}, in the 
context of a semiclassical vector model for the spin space. Hence,
in the minimal-coupling prescription of gravity, a spinning particle 
effectively probes a different geometry, lying within the general class of 
Riemann-Cartan geometry, which is determined by an effective 
metric $g^{(eff)}_{\mu\nu}$ that depends on its spin. 

In particular, in  vector models of spin the basic variables are the 
non-Grassmann vector $\omega^{\mu}$ and its conjugated momentum $\pi_{\mu}$. The 
spin-tensor is constructed using these variables as $S^{\mu \nu}=2 
\left(\omega^{\mu}\pi^{\nu}-\omega^{\nu}\pi^{\mu} \right)$. Then, one starts 
from the free theory in flat space, for which there is a Lagrangian  
formulation without auxiliary variables 
\cite{GuzmanRamirez:2013ynp,Deriglazov:2015bqa}, and the minimal coupling to 
gravity is achieved by covariantization of this   Lagrangian   
\cite{GuzmanRamirez:2013ynp,Deriglazov:2014zzm,Deriglazov:2015bqa}. After 
constructing the Hamiltonian formulation, one can eliminate the 
momenta from the Hamilton equations by using the mass-shell condition. Thus, a 
closed system for the equations of motion  (the Lagrangian form of 
MPTD equations) is obtained, with the emergence of the effective metric 
$g^{eff}_{\mu \nu}$ \cite{Deriglazov:2015bqa}.  

The effective metric, 
produced along the world-line of the particle through interaction of the spin 
with gravity, is given by  \cite{Deriglazov:2015bqa} 
\begin{eqnarray}
  &&
  \!\!\!\!\!\!\!\!\!\!\!\!\! 
  g^{(eff)} _{\mu\nu}=g_{\mu\nu}+\frac{1}{8 
m^2}\left(S^{\sigma}_{~\mu}\theta_{\sigma\nu}+S^{\sigma}_{~\nu}\theta_{\sigma\mu
}\right) \nonumber\\
&& \ \ \  \ \ 
+
\left(\frac{1}{8 m^2}\right)^2 S^{\sigma \alpha} \theta_{\sigma \mu} 
S^{\tau}_{~\alpha} 
\theta_{\tau \nu},
\label{EffMetric}
\end{eqnarray}
where     $m$ is the mass of the particle. In the above expression, the 
spin-tensor of the particle is defined as 
$S^{\mu\nu}=2 \left(\omega^{\mu}\pi^{\nu}-\omega^{\nu}\pi^{\mu} \right)=\left(S^{i 0}=D^{i}, S^{i 
j}=2 \epsilon^{i j k} S_{k}\right)$, where $D^{i}$ is the dipole electric moment 
and $S^{i}$ is the three-dimensional spin-vector 
\cite{Barut:1980aj,Deriglazov:2014tsa}, and it satisfies the relation 
$S_{\mu\nu} S^{\mu\nu}\equiv8\sigma=const.$ with $\sigma$   the absolute spin 
which is a constant of motion. Finally, the tensor 
\begin{equation}
 \theta_{\mu\nu}\equiv 
R_{\alpha \beta \mu\nu} S^{\alpha \beta},
\end{equation}
where $R_{\alpha \beta \mu\nu}$ is 
the    Riemman tensor   related to the physical 
metric $g_{\mu \nu}$, quantifies  the coupling between spin and gravity.

In the present manuscript  we desire to investigate the implications of this 
 effective spin-dependent metric in the 
context of cosmology, by deriving the corresponding modified 
Friedmann  equations. In particular, we identify the extra
spin-gravity coupling terms  as   an effective dark-energy sector. 
We mention here that since the spin-gravity coupling above is based on the 
coupling of Riemann tensor to spin, one expects 
that is would be larger  in the early Universe, or around   black 
holes. Nevertheless, since the coupling is present, even at late-times it can 
play a role if it generates a collective effect from all spining dark-matter 
particles of the Universe. In some sense the situation is similar to modified 
gravity, where the modification at late-times (where curvature is small) is 
extremely small, and thus it 
is impossible to be observed in Solar System experiments or in scales below 
galaxy clusters, however, in the whole Universe  collectively, it can lead to 
deviations from $\Lambda$CDM paradigm that can improve cosmological behavior. 
Finally, note that concerning dark matter there are many theories which 
suggest that it could correspond to  massive higher spin particles, typically 
found in string theory, and this could lead to enhanced  spin-gravity 
coupling   effects \cite{Addazi:2021xuf}.

The plan of the paper is the following: In Section \ref{model} we present the 
construction at hand, extracting the modified Friedmann equations and the 
effective dark energy sector. In Section \ref{cosmbeh} we investigate the 
resulting cosmological behavior, performing a dynamical system analysis and 
elaborating the model numerically. Finally, in Section \ref{Conclusions} we 
summarize the obtained results. Throughout the manuscript, we adopt natural 
units $c=\hbar=1$ and we use the metric 
signature $(-,+,+,+)$.

\section{Modified Friedmann equations and effective dark energy}
\label{model}

In this section we apply the above formulation in a cosmological framework, 
namely we consider  the background metric $g_{\mu \nu}$  to be 
 a flat Friedmann-Robertson-Walker (FRW) one, with form
\begin{equation}
ds^2=-dt^2+a(t)^2 \delta_{i j} dx^{i}dx^{j},
\label{FRW}  
\end{equation} 
with $a(t)$   the scale factor.
Concerning the matter sector, we consider baryonic and dark matter particles 
corresponding to the standard perfect-fluid energy-momentum tensor 
\be
T_{\mu\nu}=\left[\left(\rho_{m}+p_{m}\right)U_{\mu}U_{\nu}+g_{\mu\nu}p_{m}\right],
\label{PFluid}
\ee 
with  $\rho_{m}$ and $p_{m}$ the energy density and pressure 
respectively, while the 
four-velocity of the fluid is $U_{\mu}=(-1,0,0,0)$ such that $U_{\mu} U^{\mu} 
=-1$. Finally, concerning the field equations we consider the ones of standard 
general relativity, namely
\be
G_{\mu\nu}\equiv R_{\mu\nu}-\frac{1}{2} g_{\mu\nu} R+\Lambda 
g_{\mu\nu}=\kappa^2 
T_{\mu\nu},
\label{EinsteinE} 
\ee
where $G_{\mu\nu}$ is the Einstein tensor, $\kappa^2=8\pi G$ is the 
gravitational constant, and $\Lambda$ is the cosmological constant.

As we mentioned in the Introduction, due to the spin-gravity coupling the 
matter particles feel the effective metric  $g^{(eff)}_{\mu\nu}$ of 
\eqref{EffMetric}.  In order to calculate it one starts with the calculation of 
the  averaged effective metric $\avg{g^{(eff)}_{\mu\nu}}$ 
\cite{Gasperini:1986mv}. The volume element contains a large number of particles 
with a randomly oriented spin distribution. Several different authors have studied the 
effects at cosmological scales of matter distributions that locally contain a 
large number of randomly oriented spin particles, 
 and they have found that the microscopic 
gravitational field equations can assume a pseudo-Einsteinian form that 
includes 
spin corrections terms \cite{Hehl:1974cn,Gasperini:1986mv}. Definitely, one 
needs an averaging procedure  for these fluctuating terms 
in the microscopic domain. This is similar to 
what is done when obtaining the macroscopic Maxwell equations.

The above averaging procedure typically leads to zero spin 
average and zero spin gradient, however to non-zero average for the spin-squared 
terms  arising in 
the field equations of the theories with spin-gravity couplings
\cite{Hehl:1974cn}.
 Therefore, the averaged 
effective metric   is obtained by substituting 
\eqref{FRW} into \eqref{EffMetric} and then averaging over all possible 
directions of the three-dimensional spin-vector $\vec{S}(t)$ and the dipole 
electric moment $\vec{D}(t)$, namely $\avg{\vec{S}\cdot\vec{S}}=\vec{S}^2$, 
$\avg{\vec{D}\cdot\vec{D}}=\vec{D}^2$ and $\avg{\vec{S}}=\avg{\vec{D}}=0$. 
In the flat FRW metric, under the assumption that the 
absolute spin $S_{\mu\nu} S^{\mu\nu}\equiv8\sigma=const.$,
we have the  relations 
$ \vec{S}(t)^2/m^2=3\alpha/(4 a(t)^4)$ and 
$\vec{D}(t)^2/m^2=6\beta/a(t)^2$, where   $\alpha$ and 
$\beta$ are constants with dimensions 
of $mass^{-2}$, and thus we find  $8\sigma=\avg{S_{\mu \nu} S^{\mu \nu}}= 6 m^2 
(\alpha 
-2 \beta ) $   \cite{Gasperini:1986mv}.
Hence, we 
finally result to the averaged effective metric $\mathcal{G}_{\mu\nu}\equiv 
\avg{g^{(eff)}_{\mu\nu}}$ given by
\begin{eqnarray}
ds^2=(-1+F_{1})dt^2+a\left( t\right)^{2}\left(1+F_{2}\right)\delta_{i j} dx^{i}dx^{j},
\label{avgg}
\end{eqnarray} where
\begin{equation}
\label{F1} \!\!\!\!
F_{1}=3 \beta 
\left(\dot{H}+{{H}^{2}}\right)\left[1+\frac{\left(5\alpha-19\beta\right)
\left(\dot{H}+H^{2}\right)}{20}\right],
\end{equation} 
\begin{eqnarray}
&& 
\!\!\!\!\!\!\!\!\!\!\!\!\!\!\!\!\!\!\!\!\!
F_{2}=\beta\dot{H}\left[\frac{19\beta\dot{H} +2 \left( 
19\beta-5\alpha\right){{H}^{2}}}{20}-1\right]\nonumber\\
&&\!\!\!\!\!\!\! \! +
\frac{\left(38{{\beta}^{2}}-30\alpha\beta+19{{\alpha}^{2}}\right){H}^{4}}{40}
+\left(\alpha-\beta\right)H^{2},
\label{F2}
\end{eqnarray} 
with $H\equiv \dot{a}/a$   the Hubble function and with dots  
denoting time derivatives.
 Hence, in the case where the dipole electric moment is 
absent we have $\beta=0$, and thus $F_1=0$.
 
By substituting the averaged effective metric \eqref{avgg} into 
 the field  equations \eqref{EinsteinE}, we finally obtain the modified 
Friedmann equations 
\begin{equation}
\label{FE00}  
\frac{\dot{F_{2}}\left[\dot{F}_{2}+4\left(1+F_{2}\right)H\right]}{4 
{{\left(1+F_{2}\right) }^{2}}\left(1-F_{1}\right)}
+\frac{H^{2}}{\left(1-F_{1}\right)}=\frac{\left({{\kappa}^{2}}\rho_{m}
+\Lambda\right)}{3},
\end{equation} 
\begin{eqnarray}
&&\!\!\!\!\!\!\!\!\!\!\!\!\!\!\!\!\!\!\!\!\!
\frac{\ddot{F}_{2}}{\left(1+F_{2}\right)}-\frac{\dot{F}_{2}^2}{4\left(1+F_{2} 
\right)^{2}
}+\frac{\left[\dot{F}_{1}+6\left(1-F_{1}\right)H\right]\dot{F}_{2}}{2\left(1-F_{
1}\right){{ \left(1+F_{2}\right)}}}\nonumber\\
&& 
\!\!\!\!\!\!\!\!\!\!\!\!\!\!\!\!\!\!\!\!\!
+\frac{H 
\dot{F}_{1}}{1-F_{1}}+2\dot{H}+3 H^2=\left(\Lambda-\kappa^2 
p_{m}\right)\left(1-F_{1}\right).
\label{FEii}  
\end{eqnarray}
Note that in the case where the spin-gravity coupling switches off (in the 
spinless limit, or for a  Minkowski 
spacetime, where   $\theta_{\mu \nu}=0$), 
namely when   $\alpha$ and $\beta$ become zero, we 
obtain that $F_{1}=F_{2}=0$ and 
thus we recover standard cosmology. 
Hence, although  spinless particles follow  
geodesics related to the metric $g_{\mu \nu}$, in theories 
with spin-gravity coupling, such as the Einstein-Cartan(-Sciama-Kibble) theory, 
spinning particles follow non-geodesic paths when moving on 
gravitational fields \cite{Hojman:1978wk, 
Hojman:1976kn,Mathisson:1937zz,Papapetrou:1951pa}.   The study of spinning 
particles in general relativity 
is an old subject 
\cite{Mathisson:1937zz,Papapetrou:1951pa,Tulczyjew:1959,Dixon:1964} and 
currently the theory is under detailed investigation by many groups.

The Friedmann equations (\ref{FE00}),(\ref{FEii}) can be rewritten in the 
standard form 
\begin{eqnarray}
  &&3H^2=\kappa^2\left(\rho_{de}+\rho_m\right), \label{SFr1} \\
  &&-2\dot{H} =\kappa^2\left(\rho_{m}+p_{m}+\rho_{de}+p_{de}\right), 
\label{SFr2}
\end{eqnarray}
where we have defined an    effective 
dark energy sector  with energy density and pressure
\begin{equation}
 \label{rhode}
 \!\!\! \kappa^2 {\rho}_{{de}}=\frac{3 
F_{1}H^2}{\left(F_{1}-1\right)}+\Lambda+\frac{3\left[\dot{F}_{2}+4\left(1+ 
F_{2}\right) H 
\right]\dot{F}_{2}}{4 \left(1+F_{2}\right)^2 \left(F_{1}-1\right) },
\end{equation}
and
\begin{eqnarray}
&&
\!\!\!\!\!\!\!\!\!\!\!\!\!\!\!\!\!\!\!\!\!\!\!\!
\kappa^2 {{p}_{de}}=-\frac{F_{1} 
\left( 2\dot{H}+3 
H^{2}\right)}{\left(F_{1}-1\right)}+
\frac{{{\dot{F}_{2}}^{2}}-4\left(1+F_{2}\right)\ddot{F}_{2}}{4\left(F_{1}
-1\right)\left(1+F_{2}\right)^2 }-\Lambda\nonumber\\
&& \!\!+
\frac{1}{2{{(1\!-\!F_1)}^2}}
\!\left\{\!\frac{[\dot{F}_{1}+6(1\!-\!F_{1})H]\dot{F
}_ {2 }} { 1+F_{2}}+2 
H \dot{F}_{1}\!\right\}\! .
\label{pde}
\end{eqnarray}
Furthermore, we can introduce the 
effective dark-energy equation-of-state parameter as
\begin{equation}
w_{de}\equiv\frac{p_{de}}{\rho _{de}}.
\label{wDE1}
\end{equation}

In the averaged effective metric  \eqref{avgg} 
the matter perfect fluid \eqref{PFluid}  acquires a modified four-velocity 
vector 
$U_{\mu}=\left(-\sqrt{1-F_{1}},0,0,0\right)$, such that 
$U_{\mu}U^{\mu}=-1$. Thus,  the energy-momentum tensor for the spinning 
matter particles is  
\be
T^{\mu \nu}=\left(\rho_{m}+p_{m}\right)U^{\mu}U^{\nu}+\mathcal{G}^{\mu \nu} 
p_{m},
\ee where $\mathcal{G}_{\mu \nu}$ is the averaged effective metric 
(\ref{avgg}).
From 
the conservation law 
$
\nabla_{\mu}{T^{\mu \nu}}=0$,
where the covariant derivative $\nabla_{\mu}$ satisfies the metric 
compatibility condition $\nabla_{\mu}{\mathcal{G}^{\mu \nu}}=0$, we obtain the 
continuity equation 
\be
U^{\mu}\nabla_{\mu}{\rho_{m}}=-\left(\rho_{m}+p_{m}\right)\nabla_{\mu}{U^{\mu}},
\ee while the perfect fluid equation of motion is
\be
U^{\mu} 
\nabla_{\mu}{U^{\nu}}=\frac{1}{\rho_{m}+p_{m}}\left(U^{\mu}\nabla_{\mu}{p_{m}} 
U^{\nu}+\nabla^{\nu}{p_{m}}\right).
\ee  
Hence, in terms of the physical metric $g_{\mu \nu}$ we obtain 
  the perfect fluid equation  as
\be
U^{0}\dot{U}^{0}=\frac{1}{2}\left(\frac{\dot{F}_{1}}{1-F_{1}}\right)\left(U^{0}
\right)^2, 
\ee with $U^{0}=1/\sqrt{1-F_{1}}$. Therefore, in the case $F_{1}\neq0$ and   
$F_{1}\neq const$, we do not obtain $U^{0}\dot{U}^{0}=0$ which is the standard 
geodesic motion, as we mentioned above.

 Hence, the matter conservation
equation in the effective metric is written as
\be
\dot{\rho}_{m}+3H\left(\rho_{m}+p_{m}\right)=Q,
\label{CEq}
\ee where 
\be
Q=-\frac{3 \left(\rho_{m}+p_{m}\right) 
\dot{F}_{2}}{2\left(1+F_{2}\right)},
\ee
and this can be also cross-checked by differentiating  (\ref{FE00}) 
and inserting into (\ref{FEii}).
 Consequently,  $\rho_{de}$ and $p_{de}$ obey the evolution equation 
\be
\dot{\rho}_{de}+3H(\rho_{de}+p_{de})=-Q.
\ee
 Interestingly enough, the spin-gravity coupling gives rise to an effective 
interaction between the matter and dark energy sectors, and thus the present 
scenario exhibits the advantages of interacting cosmology 
\cite{Bolotin:2013jpa,Wang:2016lxa, Chen:2008ft, Basilakos:2008ae,
	Yang:2014hea,  Faraoni:2014vra, 
  Boehmer:2015kta,Nunes:2016dlj, 
	Mukherjee:2016shl, Pan:2017ent,  vonMarttens:2018iav, Konitopoulos:2021eav,Otalora:2013tba,Lepe:2017yvs}. As expected,  in the case where the 
spin-gravity coupling switches off we have $Q=0$ and 
thus we recover standard, non-interacting, cosmology.

We mention here that the physical metric is   $g_{\mu 
\nu}$ (for instance photons follow  null geodesics in terms of $g_{\mu 
\nu}$ and thus the CMB, SNIa etc physics is done with $g_{\mu 
\nu}$). The basic idea of the class of theories with
spin-gravity couplings is that the coupling makes dark-matter particles feel 
the effective metric $g^{(eff)}_{\mu\nu}$, and their interaction with 
$g^{(eff)}_{\mu\nu}$, namely the 
spin-gravity interaction,   from the point of view of the physical metric 
$g_{\mu \nu}$ appears as extra terms in the standard Friedmann equations (i.e. 
the Friedmann equations in terms of the physical metric), which in our model 
  provides the effective dark-energy sector. Hence, one cannot perform a 
coordinate transformation to the averaged effective metric (\ref{avgg}) and 
bring it to the standard FRW form, since this transformation freedom has 
already be spent in order to bring   $g_{\mu 
\nu}$ to the the standard FRW form.

Lastly, concerning cosmological investigations it proves convenient to introduce the 
total equation-of-state parameter as 
\be
w_{tot}=\frac{p_{de}+p_m}{\rho _{de}+\rho _m},
\label{wtot}
\ee
which is   related to the deceleration parameter $q$ through 
\be
q\equiv -1-\frac{\dot{H}}{H^2}=\frac{1}
{2}\left(1+3w_{tot}\right),
\label{deccelparam}
\ee 
as well as
 the   density 
parameters  
\be 
\Omega_{m}\equiv\frac{\kappa^2 \rho_{m}}{3 H^2}, \:\:\: \text{and} \:\:\: 
\Omega_{de}\equiv\frac{\kappa^2 \rho_{de}}{3 H^2}.
\ee 

\section{Cosmological behavior}\label{cosmbeh}

In this section we investigate the cosmological behavior of the scenario at 
hand,  by using the Friedmann equations \eqref{SFr1} and 
\eqref{SFr2}, alongside the   conservation equation \eqref{CEq}. 
As a starting model we focus on the case where the dipole electric moment is 
absent, namely $\beta=0$, thus from (\ref{F1})-(\ref{F2}) we obtain
\bea
&& F_{1}=0,\\
&& F_{2}=\frac{19}{40} \alpha ^2 H^4+\alpha  H^2.
\eea 
Substituting these into \eqref{rhode} and 
\eqref{pde},   we find
\begin{equation}
\!\!\!\!\!\!\!\!\!\!\!\!\!\!\!\!\!\!\!\!\!\!\!\!\!\!\!\!\!\!\!\!\!\!\!\!\!\!\!\!
\!\!\!\!\!\!\!\!\!\!\!\!\!\!\!\!\!\!
\kappa^2 \rho_{de}=\Lambda-6 \alpha  H^2 \dot{H} -3 \alpha ^2 H^2 \dot{H}^2,
\end{equation}
\begin{equation}\!\!
  \kappa^2 p_{de}=-\Lambda+6 \alpha  H^2 \dot{H}+2 \alpha  \dot{H}^2  +\frac{27 
\alpha^2 H^2 \dot{H}^2}{10}+2 \alpha  H \ddot{H},
\end{equation}
and then 
\be
w_{de}=-1+\frac{20 \alpha  H \ddot{H}+20 \alpha  \dot{H}^2-3 \alpha ^2 H^2 \dot{H}^2}{10 \left(\Lambda -3 \alpha ^2 H^2 \dot{H}^2-6 \alpha  H^2 \dot{H}\right)}.
\ee 
Hence, we can verify that
\bea
&& \dot{\rho}_{de}+3 H (\rho_{de}+p_{de})=-Q,\\
&& \dot{\rho}_{m}+3 H (\rho_{m}+p_{m})=Q,
\eea 
with
\be
Q=-3 \alpha (\rho_{m}+p_{m}) H \dot{H}.
\ee 
Therefore, in an expanding universe, for $\alpha>0$ we have $Q>0$ and thus energy is transferred from dark energy to dark matter sector \cite{CarrilloGonzalez:2017cll}.   

In order to study the cosmological dynamics of the scenario we introduce the 
following set of dimensionless variables
\be
X=\kappa  H, \:\:\: Y=-\frac{\dot{H}}{H^2},\:\:\: Z=\Omega_{m}.
\ee
Therefore, the set of cosmological equations can be rewritten in an autonomous 
form 
as
\bea
&& \frac{dX}{dN}= -X Y,\nonumber\\
&& \frac{dY}{dN}=-3 Y (w_m+1-Y) -\frac{\tilde{\Lambda} (w_m+1)}{2 \tilde{\alpha} X^4}+\nonumber\\
&& \frac{3}{20} \tilde{\alpha} (10 w_m+9) X^2 Y^2+\frac{3( w_m+1)-2 Y}{2 \tilde{\alpha} X^2},
\label{Auto}
\eea 
with $w_m\equiv p_m/\rho_m$ the matter equation-of-state parameter. Moreover, 
  we find     that
\be
Z= \left(\tilde{\alpha} X^2 Y-1\right)^2-\frac{\tilde{\Lambda }}{3 X^2},
\ee
with $\tilde{\alpha}=\alpha/\kappa^2$ and $\tilde{\Lambda}=\kappa ^2 \Lambda$. 
 Thus, the   dark-energy  density parameter   becomes
\be
\Omega_{de}=1-Z=1-\left(\tilde{\alpha} X^2 Y-1\right)^2-\frac{\tilde{\Lambda }}{3 X^2}.
\ee
Furthermore, in terms of these phase-space variables we obtain 
\be
w_{de}=w_m+ \frac{X^2 \left[3( w_m+1)-2 Y\right]}{3 \tilde{\alpha} X^4 Y \left(\tilde{\alpha} X^2 Y-2\right)-\tilde{\Lambda}},
\ee and
\be
w_{tot}=-1+\frac{2 Y}{3},
\ee 
and therefore  accelerated expansion 
occurs for $0\leq Y<1$.

We proceed by performing a dynamical system analysis, in order to extract the 
global behavior of the scenario, i.e. to investigate the late-time asymptotic 
behavior independently of the Universe initial conditions 
\cite{wainwrightellis1997,Coley:2003mj,Bahamonde:2017ize,Copeland:1997et,
 Leon:2010pu, Leon:2013qh, Fadragas:2013ina, Zonunmawia:2018xvf, Dutta:2016bbs}.
In order to extract the critical points of the autonomous system   \eqref{Auto} 
 we impose the conditions $dX/dN=dY/dN=0$, and in order to study their 
stability 
we calculate the eigenvalues of the involved perturbation matrix. We find 
the following fixed points:
\begin{itemize}
    \item  Point (a): $\left\{X_c=0, Y_c=\frac{3 (w_m+1)}{2}, Z_c=1\right\}$.\\
    This   point represents the dark-matter dominated era for which 
$\Omega_{m}=1$ and $\Omega_{de}=0$, while 
$w_{tot}=w_{de}=w_m$. 
    \item  Point (b): $\left\{X_c=\frac{\sqrt{\tilde{\Lambda}}}{\sqrt{3}}, 
Y_c=0, Z_c=0\right\}$.\\
    This fixed point corresponds to a dark-energy dominated accelerating 
solution for which $\Omega_{de}=1$ and $\Omega_m=0$, with $w_{tot}=w_{de}=-1$. 

\end{itemize}
Considering small perturbations around the critical points 
such that $X(N)=X_c+\delta{X}(N)$ and  $Y(N)=Y_c+\delta{Y}(N)$,  with 
$\delta{X}(N)\ll X_c $ and $\delta{Y}(N)\ll Y_c $, we define   the functions 
$F(X,Y)=dX/dN$ and $G(X,Y)=dY/dN$ and then we use them to construct 
the perturbation matrix 
$\mathcal{M}=\{\{\partial{F}/\partial{X},\partial{F}/\partial{Y}\},\{\partial{G} 
/\partial{X},\partial{G}/\partial{Y}\}\}|_{(X_c,Y_c)}$. Therefore we obtain:
\begin{itemize}
    \item   Point (a): The perturbation matrix is divergent and thus this 
critical point corresponds to unstable solution that describes the 
intermediate-time matter era of the Universe.
    
    \item  Point (b): The eigenvalues are given by 
    \be
   \mu_1=-\frac{3}{\tilde{\alpha} \tilde{\Lambda}},\:\:\: \mu_2=-3 (w_m+1).
    \ee 
    The above eigenvalues satisfy the stability conditions $\mu_1 <0$ and 
$\mu_2<0$ for $\tilde{\Lambda}>0$ and $\tilde{\alpha}>0$, and therefore this 
point is a stable node, i.e. a stable late-time 
solution.
\end{itemize}

Let us now perform a numerical elaboration of the cosmological equations.  We 
use the stability conditions obtained above, and the current 
observational values of the cosmological parameters \cite{Planck:2018vyg}. 
Additionally, it proves convenient  to introduce the 
redshift parameter $z=1/a-1$ (we set the present scale factor to $a=a_{0}=1$) as 
the independent variable, and for the present density parameters  at redshift 
$z=0$ we impose 
  $\Omega_{de}^{(0)}\simeq 
0.69$ and $\Omega_{m}^{(0)}\simeq 0.31$  
\cite{Planck:2018vyg}.

\begin{figure}[htbp]
	\centering	\includegraphics[width=0.45\textwidth]{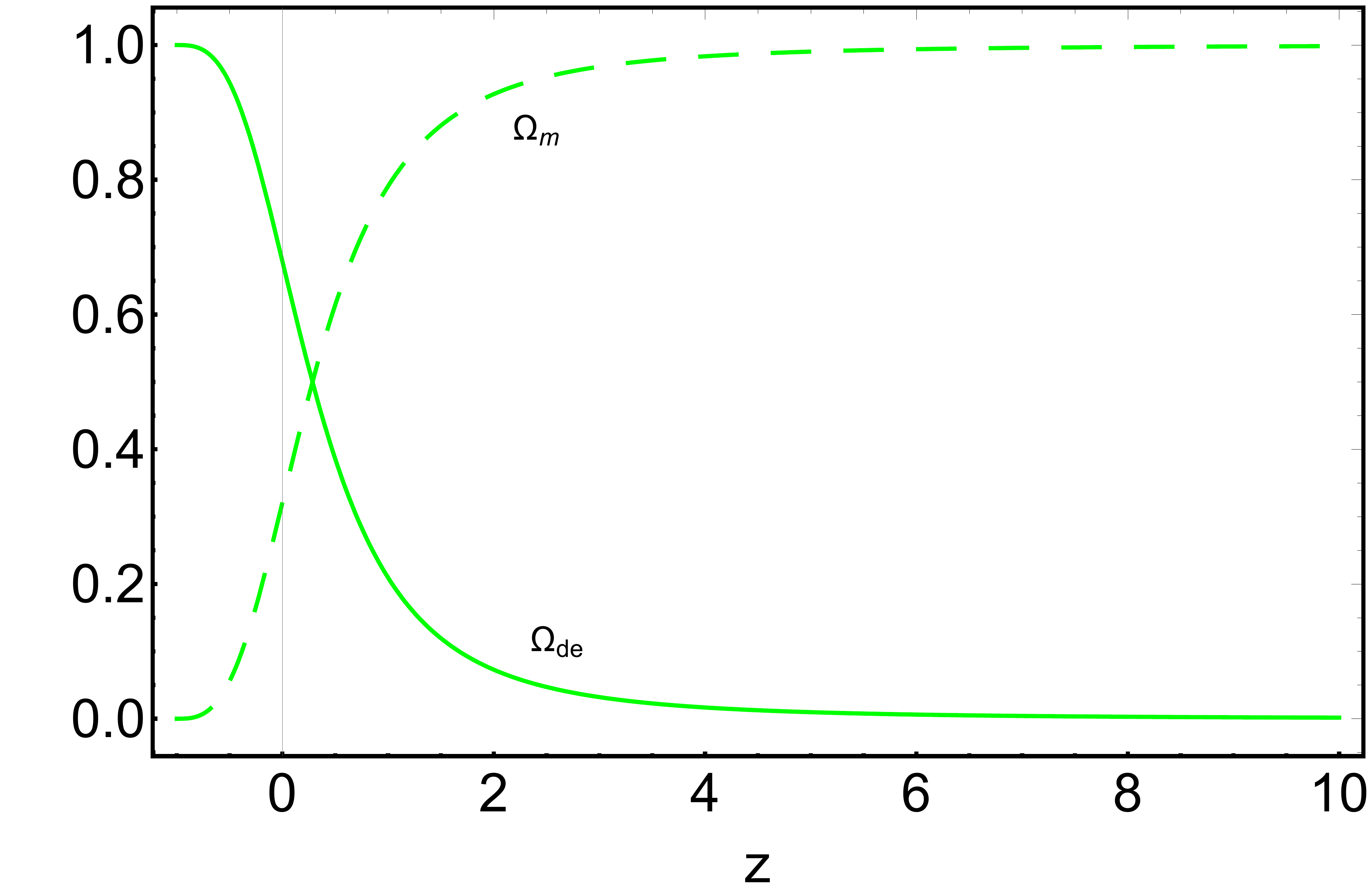}
	\caption{{\it{The matter and dark-energy density 
parameters $\Omega_{m}(z)$ and $\Omega_{de}(z)$    as functions of the 
redshift 
$z$, for $\tilde{\alpha}\equiv\alpha/\kappa^2=9\times 10^{-4}$ and $\tilde{\Lambda}\equiv \kappa^2 \Lambda=2\times 10^{-7}$ .  We have imposed 
$\Omega_{de}^{(0)}=\Omega_{de} (z=0)\simeq 0.69$ and 
$\Omega_{m}^{(0)}=\Omega_{m} (z=0)\simeq 0.31$ in agreement with observations 
\cite{Planck:2018vyg}.  }}
}  
	\label{FIG1}
\end{figure}

\begin{figure}[ht]
	\centering	
\includegraphics[width=0.45\textwidth]{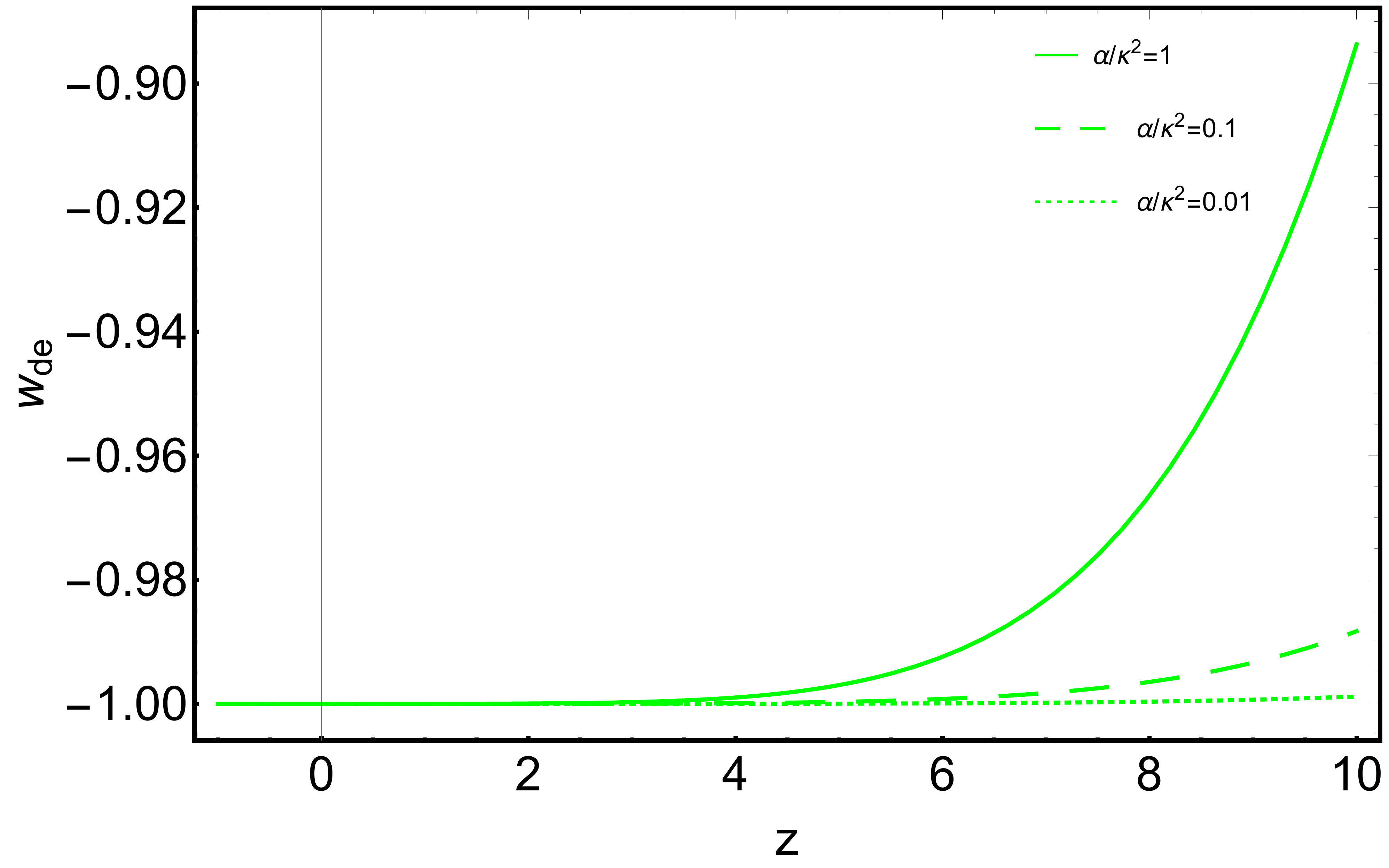}
	\caption{{\it{The 
dark-energy equation-of-state parameter $w_{de}$ as  function of the redshift 
$z$, for  $\tilde{\alpha}=0.01$ (dotted curve),  $\tilde{\alpha}=0.1$ (dashed 
curve) and  $\tilde{\alpha}=1$ (solid curve). We have imposed 
$\Omega_{de}^{(0)}=\Omega_{de} (z=0)\simeq 0.69$ and 
$\Omega_{m}^{(0)}=\Omega_{m} (z=0)\simeq 0.31$ in agreement with observations 
\cite{Planck:2018vyg}.   }  }}
	\label{FIG2}
\end{figure}

In Fig. \ref{FIG1} we depict the behavior of the matter and dark-energy density 
parameters  as functions of the redshift, and as we can see we obtain the 
expected thermal history, with the successive matter and dark-energy epochs. 
Additionally,  in Fig. \ref{FIG2} we present the behavior of the effective 
dark-energy equation-of-state parameter $w_{de}(z)$ for different values of the 
parameter $\tilde{\alpha}\equiv\alpha/\kappa^2$. From this figure one can 
clearly see the effect of the spin-gravity coupling. In particular, while at 
present times, as well as in the future ($z\rightarrow -1$), $w_{de}$ 
stabilizes at the cosmological constant value $-1$, at earlier times it 
deviates from this value, and thus the evolution of the present scenario 
deviates from that of $\Lambda$CDM concordance model.
 
\begin{figure}[htbp]
	\centering	\includegraphics[width=0.45\textwidth]{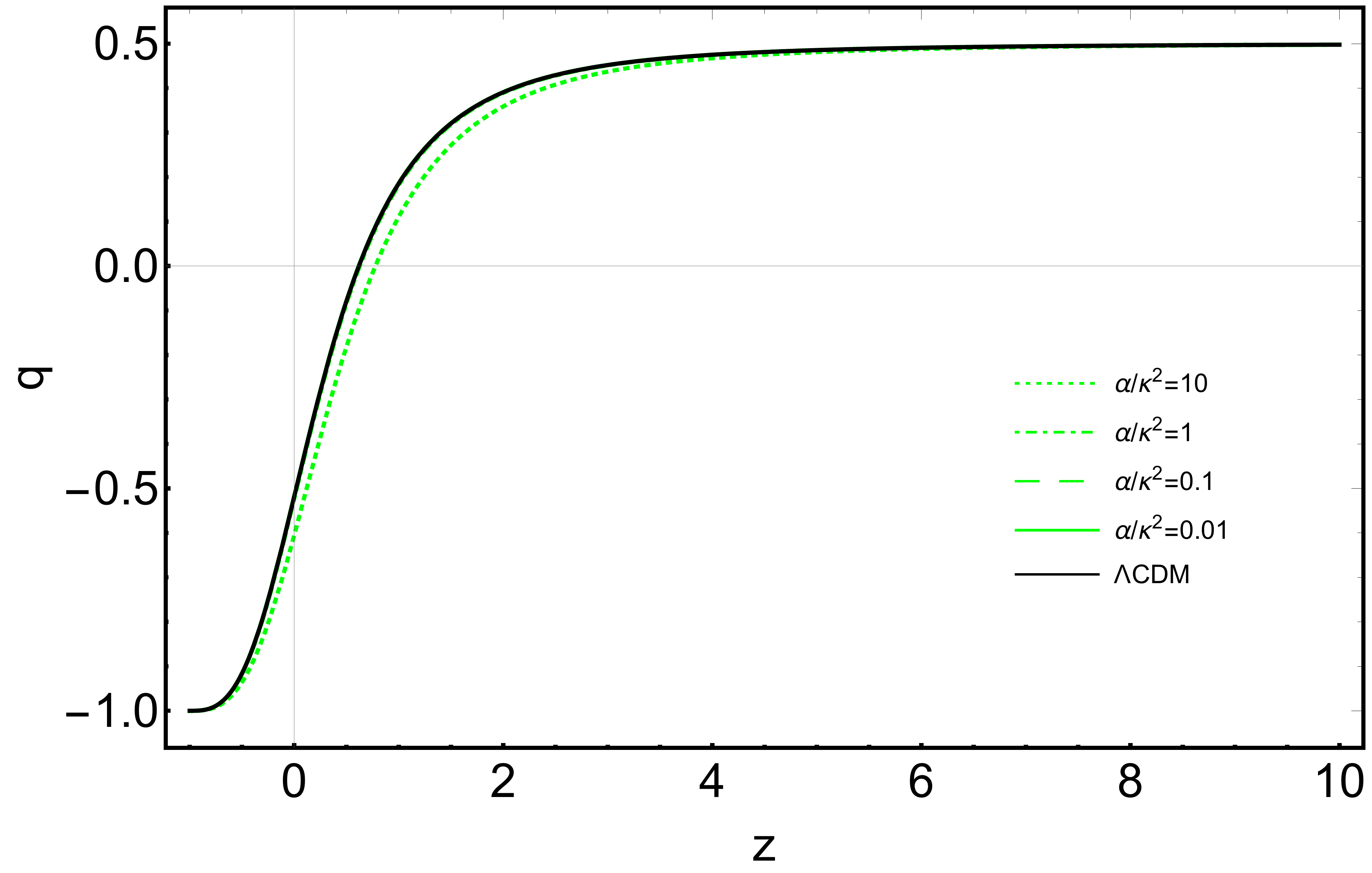}
	\caption{{\it{The deceleration parameter $q$ 
as a function of the redshift $z$, for $\tilde{\alpha}=10$ (dotted 
curve), $\tilde{\alpha}=1$ (dot-dashed 
curve), $\tilde{\alpha}=0.1$ (dashed curve) 
and  $\tilde{\alpha}=0.01$ (solid curve), as well as for the $\Lambda$CDM  
scenario. We have imposed 
$\Omega_{de}^{(0)}=\Omega_{de} (z=0)\simeq 0.69$ and 
$\Omega_{m}^{(0)}=\Omega_{m} (z=0)\simeq 0.31$ in agreement with observations 
\cite{Planck:2018vyg}.  
	}  }}
	\label{FIG3}
\end{figure}

\begin{figure}[ht]
	\centering	\includegraphics[width=0.45\textwidth]{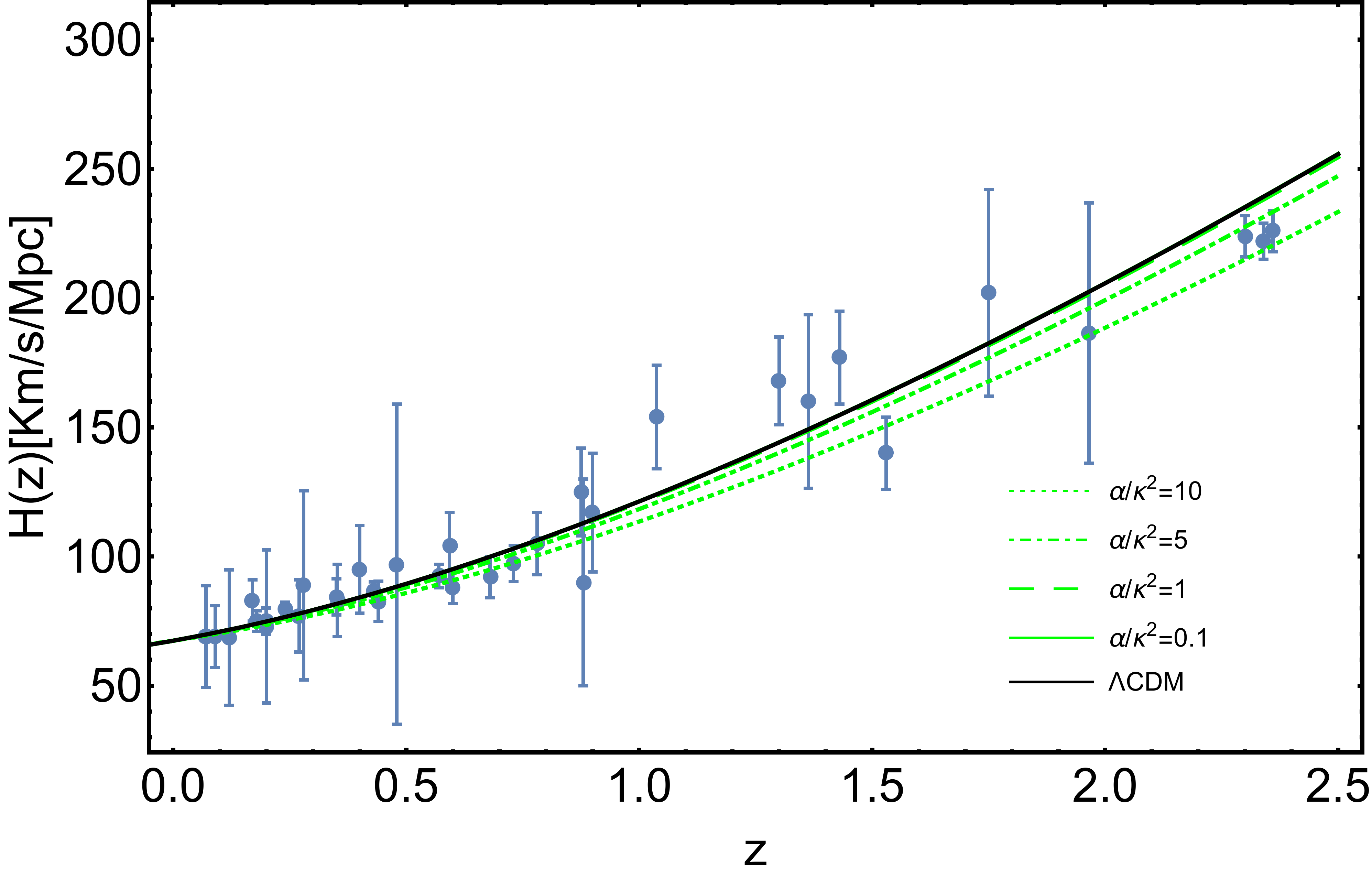}
	\caption{{\it{The Hubble function $H(z)$ as a function of the redshift $z$ 
 for     $\tilde{\alpha}=10$ (dotted 
curve),  $\tilde{\alpha}=5$ (dot-dashed curve), $\tilde{\alpha}=1$ (dashed curve)
and  $\tilde{\alpha}=0.1$ (solid curve), as well as for $\Lambda$CDM  
scenario, alongside   the Hubble data points from  \cite{Meng:2015loa} and 
\cite{Farooq:2013hq}. }  }}
	\label{FIG4}
\end{figure}

In Fig. \ref{FIG3} we show how the spin-gravity coupling can affect the behavior 
of the deceleration parameter. In particular, the 
transition from the decelerated to the accelerated regime is sensitive to the 
value of the parameter $\tilde{\alpha}$. For large values of    
$\tilde{\alpha}$ (e.g. $\tilde{\alpha}\gtrsim 10$), this transition happens at 
a redshift $z\gtrsim 0.78$, long before  $z\approx 0.65$ which is suggested by  
 observations \cite{Planck:2018vyg}, while for smaller  $\tilde{\alpha}$ 
values we obtain transition redshift values that deviate  only slightly
from the $\Lambda$CDM value. Finally, in Fig. \ref{FIG4} we present the 
predictions of our model for the Hubble function. As we observe, for 
$\tilde{\alpha}\gtrsim 10$ there is a significant deviation with respect to the 
$\Lambda$CDM predictions, however for smaller $\tilde{\alpha}$ values we can 
obtained the desired deviation from $\Lambda$CDM  scenario.

\section{Conclusions}
\label{Conclusions}

We investigated cosmological scenarios with spin-gravity coupling. In particular, due to the spin of the baryonic and dark matter particles and its 
coupling to gravity, they feel an effective  spin-dependent metric, which 
can be calculated semi-classically in the   
Mathisson-Papapetrou-Tulczyjew-Dixon formalism \cite{Deriglazov:2015bqa}. Hence, the usual field 
equations give rise to modified Friedmann equations, in which the extra terms 
can be identified as an effective dark-energy sector. Additionally, we obtained 
an effective interaction between the matter and dark-energy sectors. In the 
case 
where the   spin-gravity coupling switches off, we recover standard 
$\Lambda$CDM cosmology.

We performed a dynamical system analysis in order to examine the late-time 
asymptotic behavior, independently of the specific initial conditions of the 
Universe. Transforming the equations to an autonomous form, we extracted the 
critical points and we examined their stability. We found a matter-dominated 
point that can describe the matter era, and a stable late-time solution 
corresponding to acceleration and dark-energy domination, that attracts the 
Universe at late times. As expected, for small values of the 
 spin coupling  parameter 
$\tilde{\alpha }$, deviations from $\Lambda$CDM concordance scenario are small, 
however for larger $\tilde{\alpha }$ values they can be brought to the desired 
amount, leading to different dark-energy equation-of-state parameter behavior, 
as well as to different transition redshift from acceleration to deceleration.
Additionally, we showed that the effective interaction between matter and dark 
energy sectors 
indicates that there is an energy flow  from dark energy 
to dark matter \cite{CarrilloGonzalez:2017cll}. Finally, we confronted the model predictions with Hubble 
function data.

It would be interesting to  perform a detailed observational confrontation, 
using data from  Type Ia Supernovae (SNIa), Baryon Acoustic Oscillations 
(BAO) and Cosmic Microwave 
Background (CMB) observations. Additionally, one should 
perform a detailed perturbation analysis, in order to examine whether the 
scenario at hand changes the matter clustering behavior  and thus alleviate the 
$\sigma_8$ tension.  These interesting and necessary studies  lie beyond the 
scope of the present work and are 
left for future projects.

\section*{Acknowledgments} 
GO acknowledges Dirección de Investigación, Postgrado y Transferencia 
Tecnológica de la
Universidad de Tarapacá for financial support through
Proyecto UTA Mayor. ENS would like to acknowledge the contribution of 
the COST Action CA18108 ``Quantum Gravity Phenomenology in the multi-messenger 
approach''.





\bibliography{bio}   

\end{document}